\documentclass{aa} 
\usepackage{graphicx}
\usepackage{txfonts}
\usepackage{hyperref}
\hypersetup{
    colorlinks=true,
    linkcolor=blue,
    citecolor=blue,
    urlcolor=blue}

\usepackage[titletoc,toc,page]{appendix}

\begin{document} 

   \title{Surface parameterisation and spectral synthesis \\of rapidly rotating stars\thanks{The codes and all the files required to carry out the computations
   described in this paper are available at\\ \url{https://github.com/astrobmm/fastrot-spec}}}

   \subtitle{Vega as a testbed}

   \author{Benjamín Montesinos}

   \institute{Centro de Astrobiología (CAB) CSIC-INTA, Camino Viejo del Castillo s/n, E-28692, Villanueva de la
              Cañada, Madrid, Spain\\
              \email{bmm@cab.inta-csic.es}}

   \date{Received 7 March 2024 / Accepted 23 May 2024}

% \abstract{}{}{}{}{} 
% 5 {} token are mandatory
 
  \abstract
  % context heading (optional)
  % {} leave it empty if necessary  
   {Spectral synthesis is a powerful tool with which to find the fundamental parameters of stars. 
   Models are usually restricted to single values of temperature and gravity, 
   and assume spherical symmetry. This approximation breaks down for rapidly rotating stars.}
  % aims heading (mandatory)
   {This paper presents a joint formalism to allow a computation of the stellar structure — 
   namely, the photospheric radius, $R$, the effective temperature, $T_{\rm eff}$, and gravity, 
   $g_{\rm eff}$ — as a function of the colatitude, $\theta$, for rapid rotators with 
   radiative envelopes, and a subsequent method to build the corresponding synthetic 
   spectrum.}
  % methods heading (mandatory)
   {The structure of the star is computed using a semi-analytical approach, which is easy to 
    implement from a computational point of view and which reproduces very accurately the 
    results of much more complex codes. Once $R(\theta)$, $T_{\rm eff}(\theta)$, and 
    $g_{\rm eff}(\theta)$ are computed, the suite of codes, {\sc atlas} and {\sc synthe}, by R. Kurucz are used to synthesise spectra for a mesh of cells in which the star is divided. The appropriate limb-darkening coefficients are also computed, and the final output spectrum is built for a given inclination of the rotation axis with respect to the line of sight. All the geometrical transformations required are described in detail.}
  % results heading (mandatory)
    {The combined formalism has been applied to Vega, a rapidly rotating star almost seen 
    pole-on, as a testbed. The structure reproduces the results from interferometric 
    studies and the synthetic spectrum matches the peculiar shape of the spectral 
    lines well.}
   % conclusions heading (optional), leave it empty if necessary 
   {Contexts where this formalism can be applied are outlined in the final sections.}

   \titlerunning{Surface parameterisation and spectral synthesis of rapidly rotating stars}
   \authorrunning{Benjam\'{\i}n Montesinos}

   \keywords{stars: rotation -- stars: structure --
                stars: spectra
               }

   \maketitle
%
%-------------------------------------------------------------------

\section{Introduction}

Spectral synthesis is one of the most powerful techniques to
characterise a star.  Comparing the high-resolution spectra of a
given target with synthetic models usually provides very accurate
stellar parameters. The spectroscopic analysis must be complemented
with a detailed analysis of the spectral energy distribution, built
from photometric observations, and, when feasible, with the use of
astrometric and interferometric observations.

A vast amount of work has been done in the field of spectral
synthesis: an extensive list of the main 1D-LTE codes available can be
found in the introduction of the paper by \citet{wheeler2023}. All
these codes allow us to compute spectra for a given set of parameters; in
particular, single values of the effective temperature, $T_{\rm eff}$,
and gravity, $\log g$ (other inputs, such as metallicity, [M/H], and
microturbulence are also required). In some cases — for example, {\sc synthe}
\citep{kurucz2014}, the codes can be adapted to simulate the spectrum
of a rotating star by computing individual surface intensities at
different inclinations through the atmosphere, applying the
appropriate Doppler shifts, corresponding to the projected rotation
speed, $\varv \sin i$, to the emergent spectra.

Single values of $T_{\rm eff}$ and $\log g$ in modelling a stellar
spectrum imply the underlying limitation of spherical symmetry. This
approximation breaks down for rapidly rotating stars:\footnote{To give
an idea of what ‘rapidly rotating’ means, and in anticipation of results
that can be obtained with the models presented in this paper, for a
star with $M/M_\odot$=2.0, $R/R_\odot$=2.5, $T_{\rm pole}$=9000 K, and
$\varv_{\rm eq}$=120 km s$^{-1}$, one obtains $R_{\rm eq}/R_{\rm
  pole}\!=\!1.05$ and $T_{\rm eq}\!\simeq\!8600$ K; these values are
significant enough to assess the need of considering the oblateness of
the star in this context.} in that regime, the star becomes oblate,
and all the relevant photospheric variables, in particular the radius,
temperature and gravity, become functions of the latitude, making the
problem complex both from the theoretical and computational points of
view. Examples of the departure from spherical symmetry are the
results of the works — all based on interferometric observations — by
\citet{bouchaud2020} on Altair ($\alpha$ aql, A7V), who find that $R_{\rm
  eq}/R_{\rm pole}\!\simeq\!1.282$; \citet{domiciano2014} on Achernar
($\alpha$ Eri, B6Vpe), giving a ratio $R_{\rm eq}/R_{\rm
  pole}\!\simeq\!1.352$; and \citet{monnier2012}, on Vega ($\alpha$
Lyr, A0V), for which $R_{\rm eq}/R_{\rm pole}\!\simeq\!1.13$ (Model 3
of that paper).

In consequence, the first issue to be tackled before proceeding to the
computation of a synthetic spectrum for a rotating star is that of its
structure; in particular, finding out the dependence of $R$, $T_{\rm
  eff}$, and $\log g_{\rm eff}$ with latitude; the effective gravity, 
$g_{\rm eff}$, is defined as the vector sum of the classical gravity 
and the centrifugal acceleration (see Eqn. 4 in Sect. \ref{sec:structure}). 
This area of research has a long history, whose starting point can 
be set in the pioneering works by \citet{vonzeipel1924a,vonzeipel1924b}, 
who found that in barotropic stars the energy flux is proportional to the 
local effective gravity, leading to $T_{\rm eff}\!\propto\!g_{\rm eff}^\beta$, 
with $\beta\!=\!0.25$. This is the well-known so-called von Zeipel law, 
which was modified by \citet{lucy1967}, proposing a smooth dependence, 
$\beta\!=\!0.08$, for stars with a convective envelope. $\beta$ is 
traditionally called the ‘gravity-darkening exponent’, the term ‘gravity darkening’
encompassing all the phenomena involved when the rotation of the star
is considered — polar temperature and gravity larger than the
equatorial values, polar radius shorter than the equatorial radius — is the
common terminology today. \citet{espinosa2012} discussed the relevance 
of gravity darkening and warned about the caveats posed by the dependence
of the resulting laws on the stellar atmosphere models chosen.

The review by \citet{rieutord2006a} gives a
summary of the advances in modelling rapidly rotating stars in the decades
preceding that paper. In recent years, substantial progress has been
made. In particular, concerning the work in this paper, we mention
ESTER \citep[Evolution STEllaire en Rotation,][and references
  therein]{espinosa2013,rieutord2016b}. ESTER is the first code computing, 
  in a consistent way, 2D models of fast-rotating stars, including their
large-scale flows. A semi-analytical approximation of this code is used 
in this work. 

The main goal of this paper is to provide a set of methods, described
in as much detail as possible, to carry out from scratch the structure
and computation of the synthetic spectrum for a rotating star. To our
knowledge, there is no publicly available code to carry out these
combined tasks. In particular, the prescription presented here for
computing the stellar structure has the advantage of being valid for
any rotation and is not restricted to slow rotators, in contrast with
the von Zeipel approximation (see Sect. \ref{sec:structure}). 
A good example of the utility of these tools is the work by 
\citet{lazzarotto2023}, in which the authors combine the use of synthetic
spectra and the ESTER model to carry out a photometric determination
of the inclination, rotation rate, and mass of rapidly rotating 
intermediate-mass stars.

The paper is organised as follows: In Sect. \ref{sec:geometry},
we describe how a star with an inclination angle, $i$, with respect to 
the line of sight is seen by the observer, and how to carry out the 
projection onto a 2D plane. In Sect. \ref{sec:structure}, we describe 
how to obtain the relevant parameters for a rotating star required to 
carry out the spectral synthesis. In Sect. \ref{sec:synthesis}, we describe 
how to build the synthetic spectrum of a star where $R$, $T_{\rm eff}$, 
and $\log g_{\rm eff}$ are functions of the latitude. In Sect. \ref{sec:vega}, 
we apply the whole formalism to Vega as a testbed. Sections \ref{sec:discussion} 
and \ref{sec:conclusions} include a discussion of the results and some
conclusions. Since this paper deals with formalisms of different areas of 
stellar physics — namely, geometry, structure, spectral synthesis, and 
limb darkening — we give in each section the basic information and equations, 
and direct the reader to the appropriate references.

\begin{figure*}[!ht]
\centering
\includegraphics[scale=0.45]{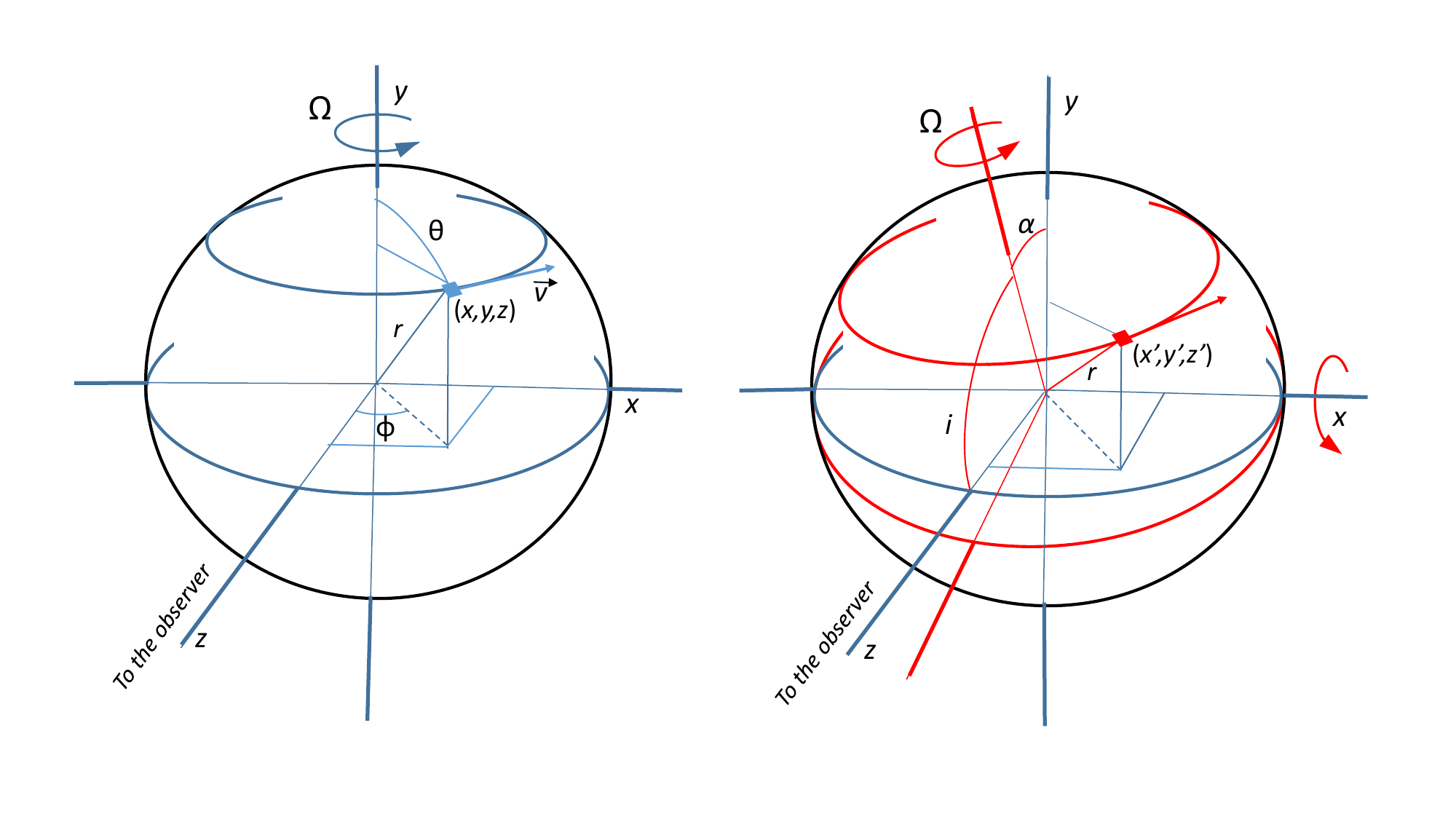}
\caption{Geometry of the problem. {\em Left}: An oblate star, rotating 
  with an angular speed, $\Omega$, and seen equator-on by the observer. 
  {\em Right}: The star is inclined by an angle, $i$, with respect to the 
  line of sight. Polar coordinates $(r,\theta,\phi)$ are used, where 
  $\theta$ is the colatitude (see Appendix \ref{app:geometry} for details).}
\label{fig:geometry_1}
\end{figure*}
\section{The geometry}
\label{sec:geometry}

Figure \ref{fig:geometry_1} shows the geometry that is used
throughout the paper. Initially, all the calculations are done
considering that the rotation axis is perpendicular to the line of
sight, which coincides with the $z$ axis (left).  The star is
then inclined by an angle of $\alpha\!=\!\pi/2\!-\!i$ around the $x$ axis,
where $i\,$ is the inclination ($i\!=\!0$, pole-on, $i\!=\!\pi/2$,
equator-on) (right).  Polar coordinates $(r,\theta,\phi)$ are
used, where $\theta$ is the colatitude (0 for latitude $\pi/2$,
$\pi$ for latitude $-\pi/2$), and $\phi$ the azimuthal angle. Since
the stellar rotation takes place around the $y$ axis, all variables
are only functions of colatitude, and are symmetric with respect to
the equator.

In Appendix \ref{app:geometry}, all the details concerning the
projection of the 3D star onto the 2D plane of the sky, and how some
quantities are seen from the point of view of the observer, are
given. Information on how to compute all the relevant geometrical
variables that arise when dealing with an oblate star is also
provided.

\section{Parameterisation of the stellar surface}
\label{sec:structure}

In this work, we follow the model of \cite{espinosa2011} (ER11, hereafter), 
also called '$\omega$-model'. A very detailed discussion of its derivation 
can be found in the work of \citet{rieutord2016a}.
Their starting point is the fact that the gravity
darkening of rapidly rotating stars is not well described by the
\citet{vonzeipel1924a} law, parameterised, as we mentioned before, as
$T_{\rm eff}\!\propto\!g_{\rm eff}^\beta$, where $T_{\rm eff}$ and
$g_{\rm eff}$ are the effective temperature and effective gravity,
respectively. Their work was triggered by the fact that some
interferometric works (see references in ER11) showed that von
Zeipel's approach seems to overestimate the temperature difference
between the pole and the equator of the star. The formalism presented
in ER11 allows the computation of $R$, $T_{\rm eff}$, and $\log g_{\rm
  eff}$ of the photosphere of a rotating star, improving the results
obtained following von Zeipel's prescription. Although simple,
mainly from a computational point of view, the ER11 model is able to
reproduce the results of more complex models; in particular, 
the above-mentioned ESTER, as can be seen in Fig. 2 of ER11. The model is
tailored for stars with radiative envelopes.

The two basic equations of the model are
\begin{equation}
\frac{1}{\omega^2 r}+\frac{1}{2} r^2\sin^2\theta=
\frac{1}{\omega^2}+\frac{1}{2}
\label{eqn:radius}
\end{equation},
\noindent that is, the Roche model (see ER11), and
\begin{equation}
\cos\vartheta+\ln\tan\frac{\vartheta}{2} = 
\frac{1}{3}\omega^2 r^3\cos^3\theta+\cos\theta+\ln\tan\frac{\theta}{2}
\label{eqn:vartheta}
,\end{equation}
\noindent where 
\begin{equation}
\omega=\Omega\sqrt{\frac{R_{\rm eq}^3}{GM}}=\frac{\Omega}{\Omega_{\rm K}}  
\label{eqn:omega}
\end{equation}
\noindent is the non-dimensional rotation rate, $\Omega$, $\Omega_{\rm
  K}$, and $R_{\rm eq}$ are the angular velocity, the Keplerian angular
velocity, and the radius, respectively, the last two at the equator, $r=R/R_{\rm eq}$,
is the non-dimensional radial coordinate, and $\theta$, as we
mentioned, is the colatitude.  $\vartheta$ is an auxiliar angular
variable (see ER11 for details, also concerning the two singularities
in Eq. (\ref{eqn:vartheta}) at $\theta=0$ and $\pi/2$).

Equation (\ref{eqn:radius}) provides the values of the photospheric
radius, $r$, as a function of $\theta$; then for each colatitude, $r$
is introduced into Eq.  (\ref{eqn:vartheta}) to obtain
$\vartheta$. Both equations can be solved by bisection, or by a
Newton-Raphson method. Once these two variables, $r(\theta$) and
$\vartheta(\theta)$ are computed, the effective gravity and the
effective temperature can be obtained from the following expressions:
\begin{equation}
\boldsymbol{g}_{\rm eff}=\left(-\frac{GM}{r^2}+\Omega^2 r \sin^2\theta\right)\boldsymbol{u}_r
+ (\Omega^2 r \sin\theta\cos\theta)\,\boldsymbol{u}_\theta
\label{eqn:geff}
\end{equation}

\begin{equation}
\begin{array}{l}
\displaystyle T_{\rm eff}\!=\!\left(\frac{L}{4\pi\sigma GM}\right)^{1/4}
                         \sqrt{\frac{\tan\vartheta}{\tan\theta}}\,g_{\rm eff}^{1/4}\\
\displaystyle \\
\displaystyle \hspace{0.5cm}=\!\left(\frac{L}{4\pi\sigma R_{\rm eq}^2}\right)^{1/4}
\left(\frac{1}{r^4}+\omega^4r^2\sin^2\theta-
\frac{2\omega^2\sin^2\theta}{r}\right)^{1/8}\!
\sqrt{\frac{\tan\vartheta}{\tan\theta}}
\end{array} 
\label{eqn:teff}
\end{equation}

We note that Eqs. (\ref{eqn:geff}) and (\ref{eqn:teff}) include three
quantities, namely, the stellar mass and luminosity, and the
equatorial radius, which — in particular $M$ and $R_{\rm eq}$ — are not
usually known with an acceptable degree of accuracy. Even the
luminosity, $L$, is a more subtle parameter to estimate in the case of
very flattened stars since the same object, seen pole-on or
equator-on, would show to the observer different spectral energy
distributions, which would lead to different apparent 
effective temperatures, and hence luminosities, the reason
being that the expression $L\!=\!4\pi\sigma\,R^2\,T_{\rm eff}^4$ loses
its meaning since both temperature and radius are functions of the
latitude.

In a practical case, when attempting to model a stellar spectrum by
building a grid of models, a reasonable range of masses, consistent with 
the estimated spectral type of the object, can be used in Eq. (\ref{eqn:geff}). 
As for the luminosity and equatorial radius, the first
parenthesis of the second expression of Eq. (\ref{eqn:teff}),
including the exponent 1/4, is basically the equatorial effective
temperature, and therefore can be substituted by an estimation of
$T_{\rm eff}^{\rm eq}$, or alternatively of the polar temperature,
$T_{\rm eff}^{\rm pole}$, using Eq. (32) of ER11:
\begin{equation}
\frac{T^{\rm eq}_{\rm eff}}{T^{\rm pole}_{\rm eff}}\!=\!
\sqrt{\frac{2}{2+\omega^2}}(1-\omega^2)^{1/12} \exp\left(-\frac{4}{3}
\frac{\omega^2}{(2+\omega^2)^3}\right)
,\end{equation}

\noindent building the grid using as inputs a range of values of
$T_{\rm eff}^{\rm eq}$ or $T_{\rm eff}^{\rm pole}$ that are also consistent
with an initial estimate of the spectral type, and iterating until
both the interferometric results are reproduced, and/or the synthetic
spectrum matches the observed one. In the most common situation, in which
interferometric observations are not available, the peculiar shape of
some spectral lines (see Sect.  \ref{sec:vega}) can give hints about
the inclination of the star, and together with an estimate of the
projected rotation speed, $\varv \sin i$, iterate using a range of
temperatures, until an agreement between the observations and the model is
reached.

We note that, according to the first expression in Eq. (\ref{eqn:teff}),
von Zeipel's law, $T_{\rm eff}\!\propto\!g_{\rm eff}^{1/4}$, is
recovered for slow rotations since as $\omega$ decreases,
$\vartheta/\theta\rightarrow 1$ (see Eq. (\ref{eqn:vartheta})). 
As a final remark, we also note that in previous works modelling
interferometric data \citep{aufdenberg2006,monnier2012} the methods
involving the calculation of the stellar radius, temperature, and
gravity make explicit use of the gravity-darkening law $T_{\rm
  eff}\!\propto\!g_{\rm eff}^\beta$, whereas the ER11 formalism used
in this work allows us to avoid the whole discussion of what the
appropriate value of the exponent, $\beta$, is.

\section{Spectral synthesis}
\label{sec:synthesis}

The spectral synthesis of a star whose relevant photospheric variables
are functions of the latitude is, from a computational point of
view, substantially more difficult than the classical
single-temperature, single-gravity modelling; however, it is
conceptually fairly intuitive and can be carried out by following these
steps:

\begin{enumerate}
    \item Once the structure is computed according to the
      prescription described in Sect. \ref{sec:structure}, the star is
      divided into cells delimited by the intersection of a mesh of
      parallels and meridians with separations of
      $\Delta\theta\!=\!\Delta\phi\!=\!1^\circ$; that implies
      $180\times360=64800$ cells, of which half are visible to
      the observer. We point out that this discretisation of the stellar 
      surface is uneven, in the sense that the areas of cells near the 
      polar regions are smaller than those of cells near the equator. 
      A discretisation keeping the surface area of all cells constant 
      \citep[see e.g.][]{bouchaud2020,lazzarotto2023} leads to exactly 
      the same results as the ones presented in this work. A finer mesh 
      does not result in any improvement or refinement of the 
      output spectrum. No numerical noise appears in the results from
      any of the discretisations.

    \item Since all variables, in particular $T_{\rm eff}$ and
      $g_{\rm eff}$, are only functions of latitude, 90 synthetic
      spectra, corresponding to the cells with colatitudes between 0
      and $\pi/2$, are computed for the corresponding values of
      temperature and gravity. The angular speed, metallicity, and
      microturbulence are fixed. These synthetic spectra, which
      contain the fluxes in erg cm$^2$ s$^{-1}$ \AA$^{-1}$, are
        not rotationally broadened.

    \item After the star is rotated by an angle, $\alpha\!=\!\pi/2\!-\!i$,
      around the $x$ axis, each individual cell is seen by the
      observer with a projected area, $(\Delta A)_{\rm p}$, a radial
      velocity, $\varv'_z$, and an angle, $\gamma$, between the line of
      sight and the normal to the surface element, the latter being
      relevant for the correction for limb darkening, $C_{\rm
        ld}(\lambda)$, to be applied (see Appendix \ref{app:limbdarkening} 
        for details of the computation of the limb-darkening coefficients). 
        Taking into account all these factors, the total flux at a given 
        wavelength, $\lambda_j$, is

    \begin{equation}
    F(\lambda_j)=\sum_{i=1}^{N_{\rm cells}} F_i(\lambda_j,\varv'_{i,\rm z})\,(\Delta A)_{i,\rm p}\,C_{\rm
    ld}(\gamma_i,\lambda_j)
    \label{eqn:flux}
    ,\end{equation}

\noindent where $F_i(\lambda_j,\varv'_{\rm z})$ is the flux at
$\lambda_j$ of the synthetic spectrum computed for the particular
values of $T_i$ and $\log g_i$ of that cell, redshifted or
blueshifted, according to the value of the radial velocity,
$\varv'_{z,i}$, of the cell (see eqns. (\ref{eqn:ap})-(\ref{eqn:mu})
and (\ref{eqn:lcdlambda})).

\end{enumerate}

The individual synthetic spectra are computed using the codes {\sc
  atlas} and {\sc synthe} \citep{kurucz2014} and the models containing
the elemental abundances and the stratification of the stellar
atmospheres as a function of temperature, gravity, metallicity, and
microturbulence velocity \citep{castelli2003}. The {\sc atlas} code
allows us to compute a model atmosphere for any value of
temperature and gravity from a close model already computed in the
Castelli \& Kurucz grids. The spectral synthesis is carried out 
using {\sc synthe}, with a resolution of $\lambda/\Delta\lambda\!=\!100\,000$ 
at 450 nm (0.0045 nm/pixel). The GNU-linux version of the codes by 
\citet{sbordone2005} is used.\footnote{The codes, models, and further 
information can be found at the URL: https://wwwuser.oats.inaf.it/fiorella.castelli/}

A grid of 3668 synthetic models with $T$ between 7000 and 20000 K
(step 100 K), $\log g$=3.0, 3.5, 4.0, 4.5, and metallicities of
[M/H]=$-2.5$, $-2.0$, $-1.5$, $-1.0$, $-0.5$, 0.0, and $+0.5$ was
computed beforehand; the microturbulent velocity is 2 km s$^{-1}$. For
all the cells at a colatitude, $\theta_i$, with parameters $(T_i,\log
g_i)$, the corresponding synthetic spectrum is computed by linear
interpolation between the four closest neighbouring models in the
grid, those bracketing at a time the temperature and the gravity of
the cell; that is, the four models in the grid — for a given metallicity —
$(T_j,\log g_k)$, $(T_{j+1},\log g_k)$, $(T_j,\log g_{k+1})$, and
$(T_{j+1},\log g_{k+1})$, have to fulfil $T_j < T_i \leq T_{j+1}$, $\log
g_j < \log g_i \leq \log g_{j+1}$.

The interpolation is easily carried out in this way: first, two
constants are defined as
\begin{equation}
C_{\rm T}=\frac{T_{j+1}-T_i}{T_{j+1}-T_j} \hspace{1cm} C_{\rm g}=\frac{\log g_{j+1}-\log g_i}{\log g_{j+1}-\log g_j}
,\end{equation}
\noindent then two intermediate models are computed, which are combined 
to give the final one $(T_i,\log g_i)$ for the $i$-th cell:
\begin{equation}
\begin{array}{l}
\displaystyle (T_i,\log g_j)\,\,\,\,=C_{\rm T}\cdot(T_j,\log g_j)\,\,\,\,\,+(1-C_{\rm T})\cdot(T_{j+1},\log g_j) \\
\displaystyle (T_i,\log g_{j+1})=C_{\rm T}\cdot(T_j,\log g_{j+1})+(1-C_{\rm T})\cdot(T_{j+1},\log g_{j+1}) \\
\displaystyle \\
\displaystyle (T_i,\log g_i)=C_{\rm g}\cdot(T_i,\log g_j) + (1-C_{\rm g})\cdot(T_i,\log g_{j+1})
\end{array}
.\end{equation}

\section{Vega as a testbed}
\label{sec:vega}

Vega ($\alpha$ Lyr, HD 172167, HIP 91262, HR 7001) is one of the most
extensively studied stars. It is well known that it is used as
standard for the calibration of several photometric systems
\citep{bessell2005} and that it is surrounded by a debris disc,
discovered by \citet{aumann1984}, which triggered an intensive study
at infrared wavelengths (\citealt{sibthorpe2010}, and references
therein).  However, this object turned out not to be the perfect
standard, showing anomalies in its luminosity
\citep{petrie1964,millward1985}, some peculiarly shaped absorption
lines \citep{gulliver1991}, and its radius \citep{hanbury1967,ciardi2001}
in comparison with other A0 V stars.

Concerning this paper, our interest focuses on the fact that all these
anomalies are now explained by the fact that Vega is a rapidly
rotating star being seen almost pole-on; that is, with a small inclination
angle, as has been shown by a number of interferometric studies
\citep{aufdenberg2006,peterson2006,monnier2012} and spectroscopic
analyses
\citep{gulliver1991,hill2004,takeda2008b,yoon2010,hill2010,takeda2021}. Table
1 in \citet{takeda2021} gives a summary of the values of the projected
equatorial velocity, $\varv_{\rm eq} \sin i$, the equatorial velocity,
$\varv_{\rm eq}$, the inclination angle, $i$, the polar and equatorial
radii, $R_{\rm pole}$ and $R_{\rm eq}$, and the rotation period, $P$,
according to different analyses.

To check the reliability of the methods described in this paper, 
we use some of the results of the above-mentioned works to check
whether the structural model (Sect. \ref{sec:vega_params}), and then
the spectral synthesis model (Sect. \ref{sec:vega_spec}), can
reproduce the observed properties.

\begin{figure}[!h]
\centering
\includegraphics[width=9.0cm]{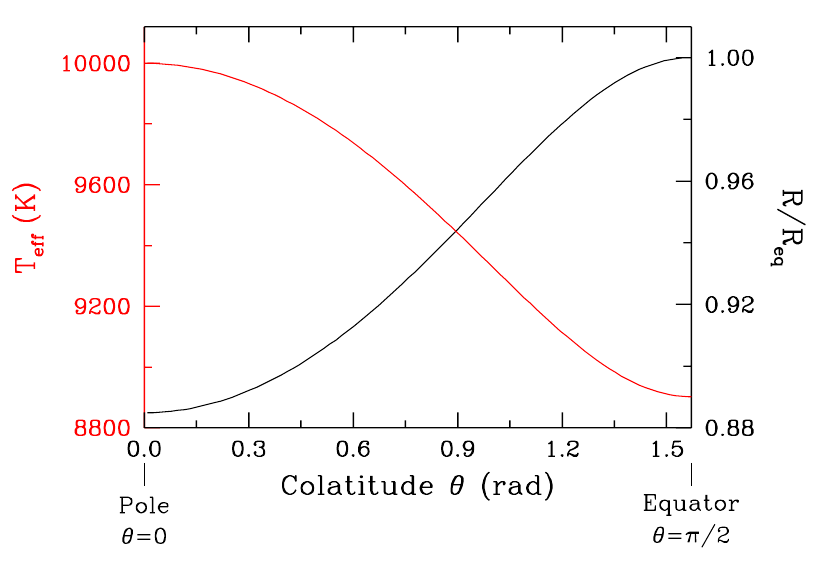}
\caption{Results for Vega from the model computed using the input
  parameters shown in Table \ref{tab:structure} and the formalism
  described in Sect.  \ref{sec:structure}. Radius, normalised to the
  equatorial radius (black), and temperature (red) are plotted
  against the colatitude, $\theta$.}
\label{fig:vega_r_t}
\end{figure}

\begin{table}[!ht]
\setlength{\tabcolsep}{9pt}  
\caption{Model structure for Vega.}
\label{tab:structure}
\begin{tabular}{lr}
\hline \hline
\noalign{\smallskip}
\multicolumn{2}{c}{Input parameters} \\
\noalign{\smallskip}\hline
\noalign{\smallskip}
Inclination, $i$ (degrees)                    & 6.2$^{\rm a}$   \\
Stellar mass, $M/M_\odot$                      & 2.15$^{\rm a}$  \\
Polar temperature, $T_{\rm pole}$ (K)           & 10000           \\
Equatorial radius, $R_{\rm eq}/R_\odot$         & 2.726$^{\rm a}$ \\
Normalised $\omega$, $\Omega/\Omega_{\rm K}$   & 0.510            \\
Metallicity, [M/H]                            & $-0.50$
\end{tabular}
\begin{tabular}{lrcr}
\noalign{\smallskip}\hline
\noalign{\smallskip}
\multicolumn{4}{c}{Derived parameters} \\
\noalign{\smallskip}\hline
\noalign{\smallskip}
                            & This work && Other works \\
\cline{2-2}\cline{4-4}\noalign{\smallskip}
$R_{\rm pole}/R_\odot$                &  2.412 && 2.418$\pm$0.012$^{\rm a}$\\
$R_{\rm eq}/R_{\rm pole}$             &  1.130 && 1.127$^{\rm a}$\\
$T_{\rm eq}$ (K)                      &  8902  && 8910$\pm$130$^{\rm a}$\\
$L_{\rm bol}/L_\odot$                 & 46.5   && 47.2$\pm$2.0$^{\rm a}$\\
$T_{\rm eq}/T_{\rm pole}$             & 0.890  && 0.885$^{\rm a}$\\
$\log g_{\rm eq}$                     & 3.769  && \\
$\log g_{\rm pole}$                   & 4.005  && \\
$\varv_{\rm eq}$ (km s$^{-1}$)        & 197.8  && 195$\pm$15$^{\rm b}$\\
$\varv_{\rm eq} \sin i$ (km s$^{-1}$) &  21.4  && 21.6$\pm$0.3$^{\rm b}$\\
Averaged $T_{\rm eff}^{\rm aver}$ (K) &  9430  && 9360$\pm$90$^{\rm a}$\\
Averaged $\log g_{\rm eff}^{\rm aver}$ & 3.958  && \\
Stellar surface area                  & 11.536$^\dagger$ &&\\
\noalign{\smallskip}
\noalign{\smallskip}\hline
\noalign{\smallskip}
\multicolumn{4}{l}{Refs.: (a) \citet{monnier2012}, (b) \citet{takeda2021}.}\\
\multicolumn{4}{l}{($\dagger$): In units of $(R/R_{\rm eq})^2$. A sphere with $R\!=\!1$ would have}\\
\multicolumn{4}{l}{a surface area $A\!=\!12.566$.}\\
\noalign{\smallskip}\hline
\end{tabular}
\end{table}

\begin{figure*}[!t]
\centering
\includegraphics[width=14.0cm]{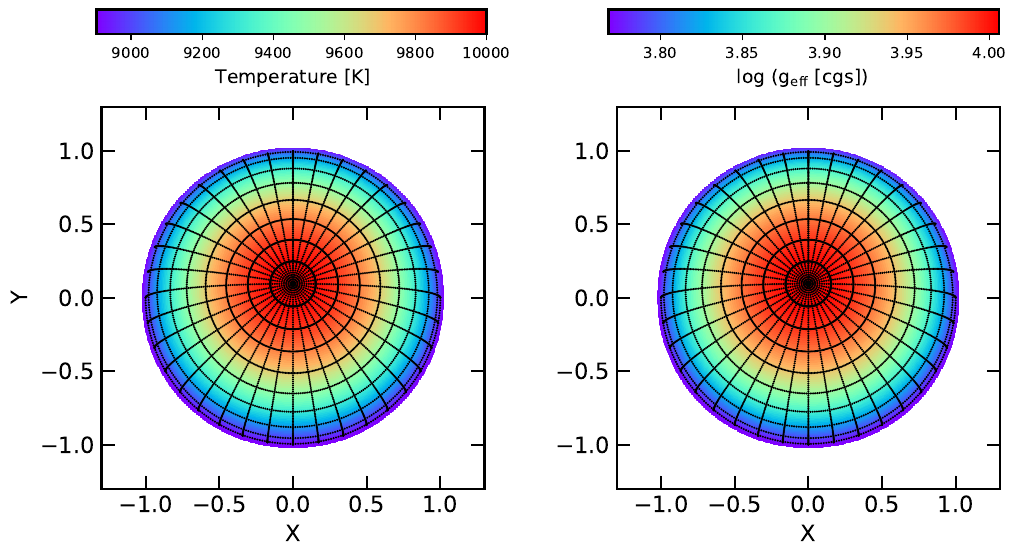}
\caption{2D structure of the temperature and effective gravity of
  Vega as seen by an observer. Parallels and meridians separated by 10
  degrees are also shown. The values in the axes are distances scaled
  to Vega's equatorial radius.}
\label{fig:tlogg}
\end{figure*}
\subsection{Structure and stellar parameters}
\label{sec:vega_params}

Table \ref{tab:structure} shows in the upper part the input parameters
of the model; the inclination, stellar mass, and equatorial radius
have been taken from \citet{monnier2012} (their Model 3, the ‘concordance
model’). Since the model also requires the polar temperature and
$\omega$ as inputs, $T_{\rm pole}$ was explored within the uncertainty 
interval given by Monnier et al. to match the luminosity, and $\omega$
was fixed to match the value of $\varv \sin i$ derived by 
\citet{takeda2008a}. The lower part of the table (Col. 2) shows the 
results of the formalisms described in Sects. \ref{sec:geometry} and 
\ref{sec:structure}; Col. 3 shows, for comparison, some parameters derived 
from interferometric \citep[Model 3 by][]{monnier2012} and spectroscopic 
analyses \citep{takeda2021}.

The results derived in this work are in general consistent with those
from previous modellings, although some discrepancies are apparent:
the temperature drop from pole to equator, 1098 K in our case, is in
agreement with that by \citet{monnier2012} (1160 K); both are
substantially smaller than those by \citet{peterson2006} ($>2400$ K),
and \citet{aufdenberg2006} (2250 K).

Some details about the calculations: the stellar luminosity was
computed by adding for all cells the quantity $\sigma (\Delta A)_i
T_{{\rm eff},i}^4$, where $(\Delta A)_i$ and $T_{{\rm eff},i}$ are the
surface area and the effective temperature of the $i$-th cell; and the
average effective temperature was estimated from the expression
$L\!=\!4\pi R_{\rm aver}^2\sigma T_{\rm eff,aver}^4$, where $R_{\rm
  aver}$ is an average of the radius, computed in the interval of
colatitudes, $[0,\pi/2]$.

Figure \ref{fig:vega_r_t} shows the stellar radius normalised to
$R_{\rm eq}$ (black) and the temperature (red), plotted against the
colatitude for the northern hemisphere (the results for the southern
hemisphere are symmetrical). Fig. \ref{fig:tlogg} shows two colour
plots showing the temperature and effective gravity profiles for Vega
according to the results of our modelling.

\subsection{Spectral synthesis}
\label{sec:vega_spec}

In this section, we check whether the results derived in
Sect. \ref{sec:vega_params} together with the formalism described in
Sect. \ref{sec:synthesis} are able to reproduce the peculiar shape of
some features of the spectrum of Vega. The high-resolution,
high-signal-to-noise spectrum atlas of Vega from \citet{takeda2007}
has been used throughout.  We do not intend here to make an analysis
of the elemental abundances, which has already been carried out by other
authors \citep[see e.g.][]{ilijic1998,qiu2001,takeda2008a,takeda2008b},
but rather to make sure that the whole set of procedures described in
the previous sections is able to reproduce the peculiar shapes of the
absorption lines of the spectrum of Vega.

\begin{figure*}[!h]
\centering
\includegraphics[width=14.0cm]{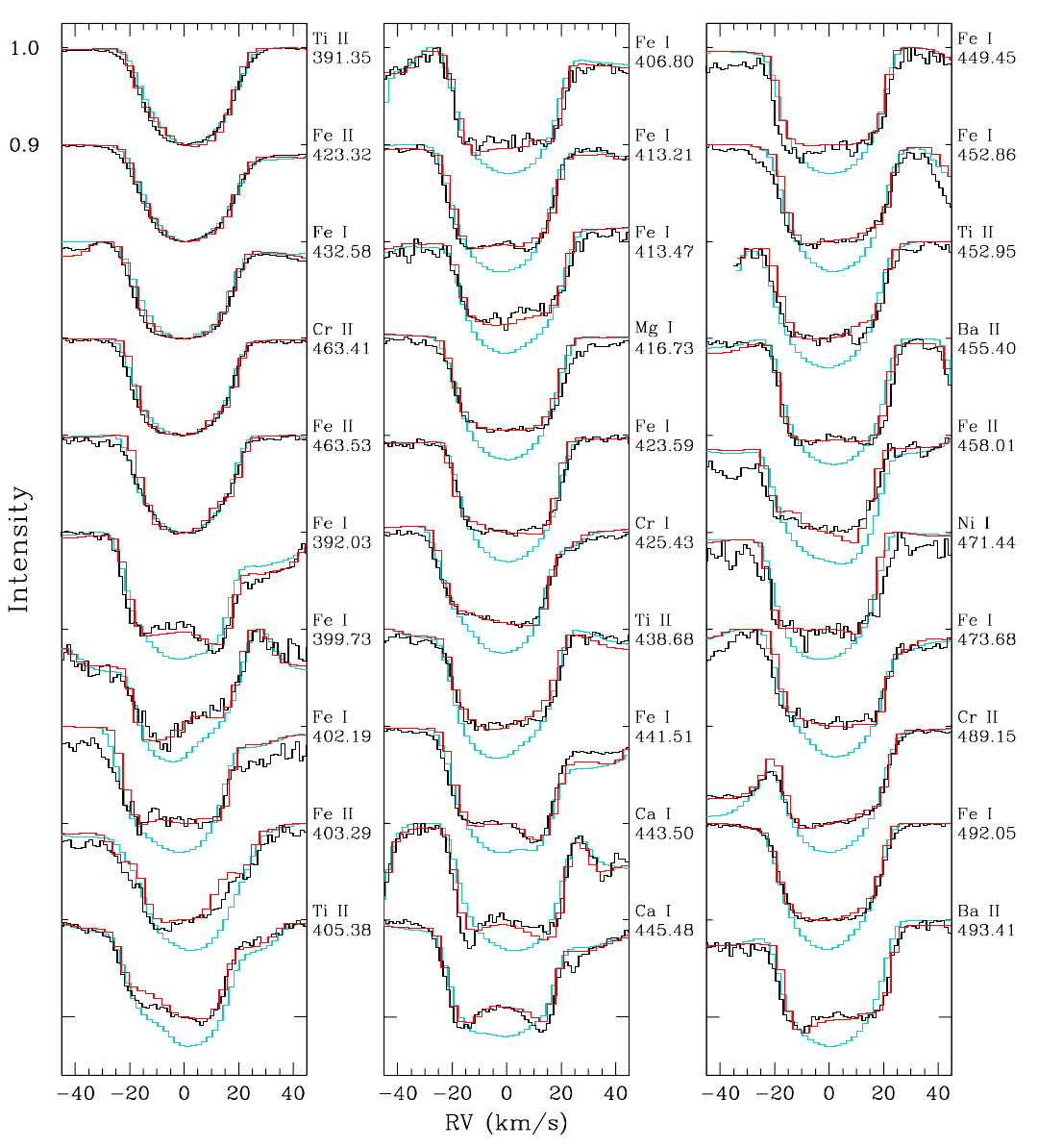}
\caption{Plotted in black, profiles of 30 absorption features of the
  spectrum of Vega from different species. Superimposed on the
  observed lines are the results of the spectral synthesis described in
  Sect. \ref{sec:synthesis}, and the spectral single-$T$, single-$\log
  g$ synthesis carried out with the values of $T_{\rm eff}^{\rm aver}$
  and $\log g_{\rm eff}^{\rm aver}$, listed in Table \ref{tab:structure}, 
  plotted in red and light blue, respectively (see the text for 
  further details).}
\label{fig:panel-lines}
\end{figure*}

Figure \ref{fig:panel-lines} shows the profiles of 30 lines, from
different species, both neutral and ionised. The observed profiles and
the results of our modelling are plotted in black and red,
respectively. In cyan, the profiles resulting from a
single-temperature, single-gravity synthetic spectrum computed with
the average parameters, $T_{\rm eff}^{\rm aver}$ and $\log g_{\rm
  eff}^{\rm aver}$, listed in Table \ref{tab:structure}, are also
plotted.  For the sake of a better graphical display of the whole set
of lines, both the observed and the synthetic profiles have been
re-scaled and normalised, placing the continuum at intensity 1.0 and
the bottom of the profiles at intensity 0.9, whereas the
single-temperature, single-gravity profiles have been scaled so that
they fit the wings of the observed absorptions.

The first five profiles of the left panel, from top to bottom, show
rounded shapes, typical of lines broadened with a classical rotation
profile; however, the remaining 25 profiles have a variety of shapes, the 
most extreme cases being those that are almost rectangular, such as Fe {\sc i}
402.19, 406.80, 449.45 nm and Ba {\sc ii} 455.40, 493.41 nm, or those
showing two deeper components at velocities close to $\pm\varv_{\rm
  eq} \sin i$, like Ca {\sc i} 445.48 nm. It is apparent that in all
cases the agreement between the shape of the observed profiles and the
results from the formalism described in Sect. \ref{sec:synthesis} is
remarkable. To our knowledge, the only work that also accurately 
reproduces the peculiar profiles of the Vega spectrum is that from
\citet{takeda2008b}.

\begin{figure*}[!ht]
\centering
\includegraphics[width=9.0cm]{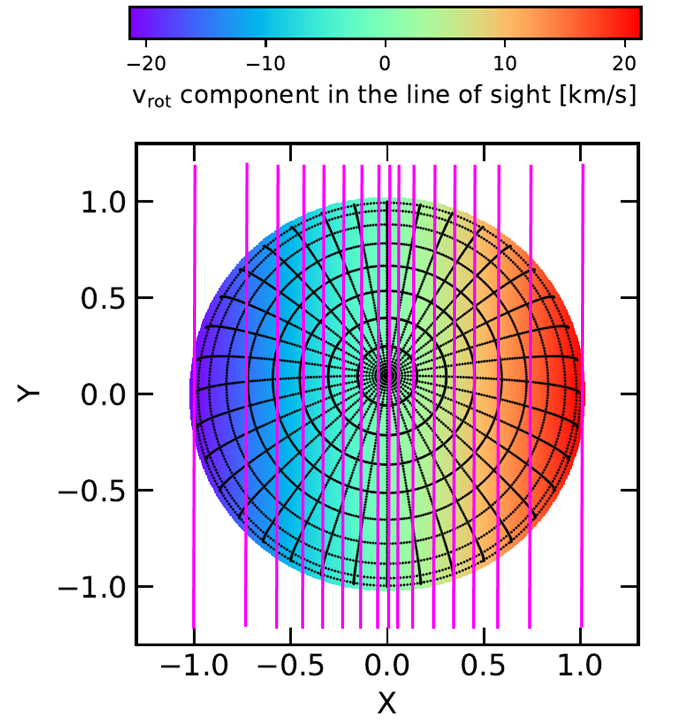}
\includegraphics[width=9.0cm]{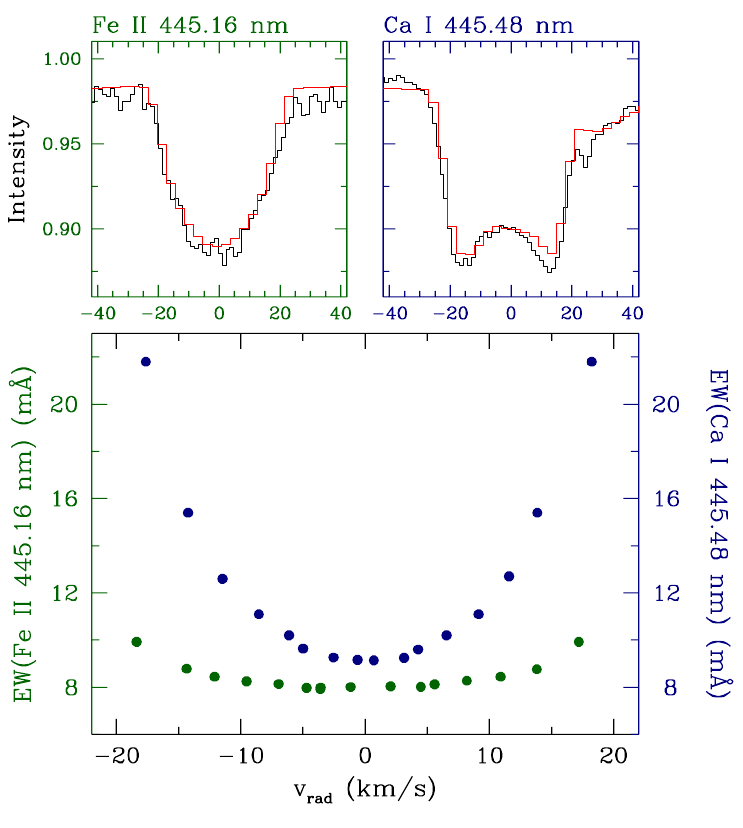}
\caption{{\em Left}: Colour plot of the projected radial velocity of
  each surface element of the star. The purple lines delimit 16
  regions contributing the same amount of flux to the continuum
  in the region near the Fe {\sc ii} 445.16 and Ca {\sc i} 445.48 nm
  lines.  {\em Right top}: Observed (black) and model (red) profiles
  of the Fe {\sc ii} (green frame) and Ca {\sc i} (blue frame)
  lines. {\em Right bottom}: Equivalent widths of the contributions to
  the lines from each of the individual 16 strips shown in the left
  panel.}
\label{fig:vstrips}
\end{figure*}

In order to understand the peculiar profile of some lines, we show as
an example an analysis of two nearby lines; namely, Fe {\sc ii} 445.16
nm, which has a rounded shape, and the above-mentioned Ca {\sc i} 445.48
nm. The left panel of Fig. \ref{fig:vstrips} shows the colour plot of
the radial velocity of each surface element for the Vega model. It is
well known that the loci of equal radial velocities, in the case of
solid-body rotation, as seen by an observer, are lines parallel to 
the rotation axis projection; that is, all points in the disc with 
$x$=constant \citep[see e.g.][chapter 18]{gray1992} have the same value of
radial velocity. Since the star is seen almost pole-on, the regions
close to the borders of the disc — the limb, which coincides with the
equator — and in particular those with the highest
projected rotation speeds, are the ones with the lowest temperatures
and gravities. Therefore, the ionisation balance of some species differs
from regions with low temperatures and gravities to regions near the
pole (see Fig. \ref{fig:tlogg}). The vertical purple lines superimposed
on the colour plot of the star in the left panel of
Fig. \ref{fig:vstrips} delimit 16 strips. Their widths have been
computed to fulfil the condition that each one contributes to the
full synthetic spectrum with the same amount of flux in the continuum
near the lines.

The right panel of Fig. \ref{fig:vstrips} shows the stellar lines
(black), with the corresponding model profiles (red) framed in green
(Fe {\sc ii}) and blue (Ca {\sc i}) boxes, respectively, and the
equivalent widths (EWs) of the contributions to the profiles from each
one of the 16 strips, using the same colour code. The different ranges
of EWs for both lines are apparent: whereas the values for the Fe {\sc
  ii} values are more even, leading to a rounded profile, the outer
strips dominate the absorption of the Ca {\sc i} line, producing its peculiar
shape.

\section{Discussion}
\label{sec:discussion}

\subsection{The $\omega$-model versus the von Zeipel approach}

Despite the fact that the $\omega$-model, and the corresponding
spectral synthesis, work well for Vega, it is interesting to point out that
\citet{takeda2008a}, using the Roche model and the von Zeipel value of the
gravity darkening exponent, $\beta\!=\!0.25$, also found a good agreement
between modelling and observations. \citet{monnier2012} showed that the
value of $\beta$ that best fit the observations was $0.231\!\pm\!0.028$,
in agreement with the von Zeipel value; therefore, both the $\omega$-model
and the von Zeipel approximation give accurate results for Vega. 

To show the real potential of the $\omega$-model, it is interesting
to explore the case of a much faster rotator for which the $\beta$ exponent is
much less then 0.25. The case of Achernar is a good one to check, since 
a value of $\beta\!=\!0.166$ must be used to reproduce the observed 
results \citep{domiciano2014}. The $\omega$-model, using as inputs — all 
extracted from Domiciano de Souza et al.'s paper — $M/M_\odot\!=\!6.1$, 
$T_{\rm pole}\!=\!17124$ K, $R_{\rm eq}/R_\odot\!=\!9.17$, and $\omega\!=\!0.838$, 
gives $T_{\rm eq}\!=\!12700$ K, in excellent agreement
with the best fit of the CHARRON RVZ model to the VLTI/PIONIER H band
observations, which gives $T_{\rm eq}\!=\!12673$ K. In contrast, the
von Zeipel model ($\beta\!=\!0.25$) gives $T_{\rm eq}\!=\!10880$ K, almost
1800 K off the value derived from observations, which is nicely 
reproduced by the $\omega$-model \citep[see][for further details]{domiciano2014}.
Obviously, that deviation would have a large impact on reproducing the 
observed spectrum, via spectral synthesis.

\subsection{Contexts in which this work can be useful}

Once the whole formalism of the $\omega$-model plus the spectral
synthesis have been put together and successfully tested with the 
paradigmatic case of Vega, and the reassuring case of Achernar mentioned 
in the previous subsection, it is interesting to point out explicitly 
in which contexts all this can be useful. With this purpose, we have 
computed several models whose details are given in Table \ref{tab:discussion}. 
The models share as fixed inputs some of the parameters of the structural 
model of Vega (see Table \ref{tab:structure}); namely, the stellar mass, 
polar temperature, equatorial radius, and metallicity. Models in the upper 
half of Table \ref{tab:discussion} have been computed with a fixed inclination,
$i\!=\!6.2^\circ$, the value obtained for Vega, and five values of
$\omega$, resulting in five values of $\varv \sin i$; namely 10, 15,
20, 21.4 (Vega), and 25 km s$^{-1}$. Models in the lower half of Table
\ref{tab:discussion} have been computed with a fixed value of the
projected $\varv \sin i$, 21.4 km s$^{-1}$, which is again the value we
obtained for the Vega model, and five values for the inclination;
namely, 5, 6.2 (Vega), 10, 15, and 20 degrees.

\begin{table}[!ht]
\caption{Parameters for the models in Figure \ref{fig:discussion}}
\label{tab:discussion}
\begin{tabular}{lr}
\hline \hline
\noalign{\smallskip}
\multicolumn{2}{l}{Fixed parameters for all models}\\
\noalign{\smallskip}
\hline
\noalign{\smallskip}
Stellar mass, $M/M_\odot$ & 2.15 \\
Polar temperature, $T_{\rm pole}$ (K) & 10000 \\
Equatorial radius, $R_{\rm eq}/R_\odot$ & 2.726 \\
Metallicity, [M/H] & $-0.5$\\
\noalign{\smallskip}
\end{tabular}
\begin{tabular}{ccccc}
\hline 
\noalign{\smallskip}
\multicolumn{5}{l}{Upper panel: inclination, $i\!=\!6.2^\circ$} \\
\noalign{\smallskip}
\hline
\noalign{\smallskip}
$\omega$ & $\varv_{\rm eq} \sin i$ & $\varv_{\rm eq}$ & $T_{\rm eq}/T_{\rm pole}$ & $R_{\rm eq}/R_{\rm pole}$ \\
         & (km s$^{-1}$) &  (km s$^{-1}$) & \\
\noalign{\smallskip}\hline
\noalign{\smallskip}
0.238 & 10.0 & 92.3  & 0.973 & 1.028 \\
0.357 & 15.0 & 138.6 & 0.942 & 1.064 \\
0.476 & 20.0 & 184.6 & 0.903 & 1.113 \\
{\em 0.510} & {\em 21.4} & {\em 197.8} & {\em 0.890} & {\em 1.130} \\
0.596 & 25.0 & 231.2 & 0.857 & 1.178 \\
\noalign{\smallskip}
\hline
\noalign{\smallskip}
\multicolumn{5}{l}{Lower panel: $\varv_{\rm eq} \sin i\!=\!21.4$ km s$^{-1}$}  \\
\noalign{\smallskip}
\hline
\noalign{\smallskip}
$i$        & $\omega$ & $\varv_{\rm eq}$ & $T_{\rm eq}/T_{\rm pole}$ & $R_{\rm eq}/R_{\rm pole}$ \\
($\circ$)  &          & (km s$^{-1}$)    &  & \\
\noalign{\smallskip}\hline
\noalign{\smallskip}
 5.0 & 0.632 & 245.2 & 0.842 & 1.200 \\
 5.0 & 0.632 & 245.2 & $\,\,\,\,\,\,\,\,0.804\,(\dagger)$ & 1.200\\
{\em 6.2} & {\em 0.510} & {\em 197.8} & {\em 0.890} & {\em 1.130} \\
10.0 & 0.318 & 123.4 & 0.953 & 1.051 \\
15.0 & 0.213 &  82.6 & 0.978 & 1.023 \\
20.0 & 0.161 &  62.5 & 0.987 & 1.013 \\
\noalign{\smallskip}
\hline
\noalign{\smallskip}
\multicolumn{5}{l}{Notes: The models in italics correspond to Vega.} \\
\multicolumn{5}{l}{($\dagger$) Model computed under the von Zeipel approximation.}\\
\noalign{\smallskip}
\hline
\end{tabular}
\end{table}

\begin{figure}[!h]
\centering
\includegraphics[width=9.0cm]{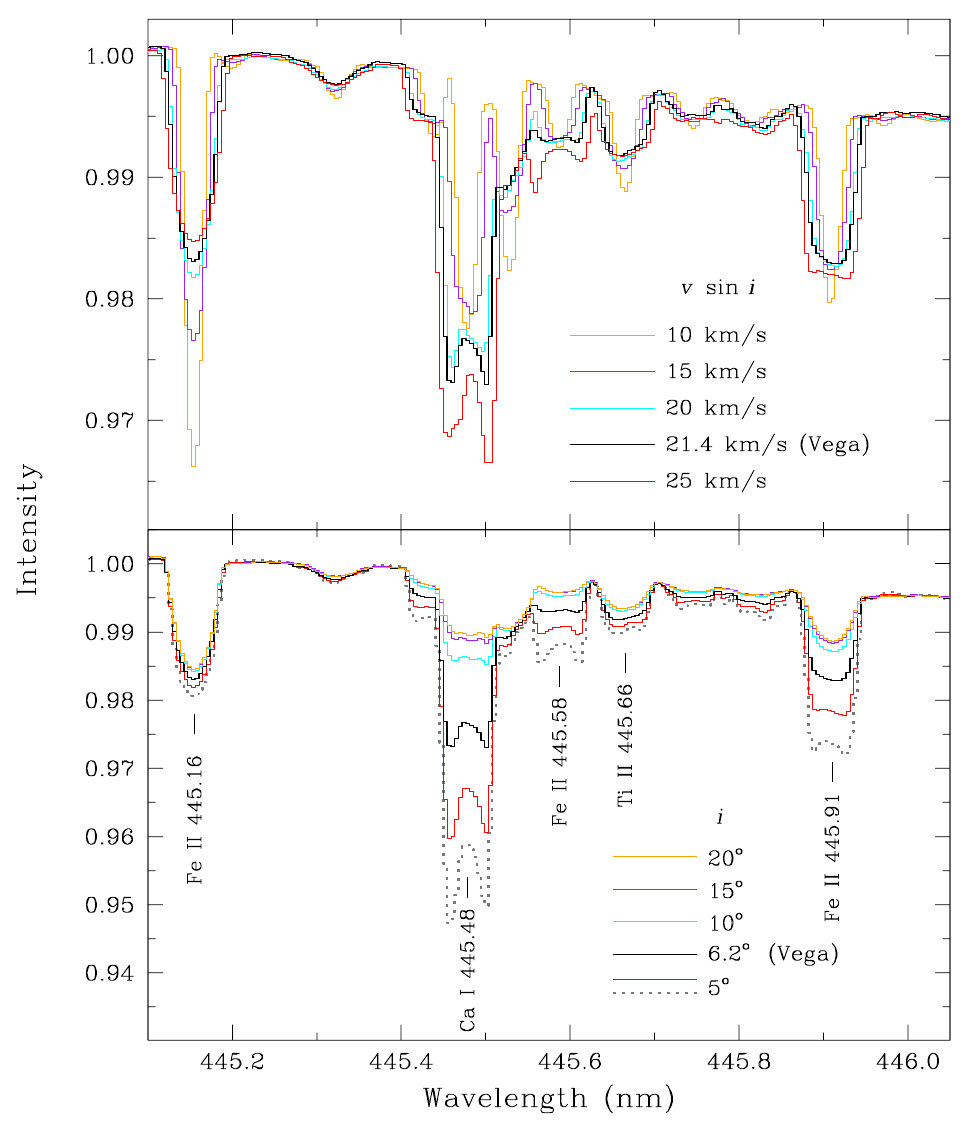}
\caption{Models of a small section of the optical spectrum showing the
  variations in the spectral lines with $\varv_{\rm eq} \sin i$ for a
  fixed inclination (upper panel), and with the inclination, $i$, for
  a fixed value of $\varv_{\rm eq} \sin i$. The parameters of the
  models are given in Table \ref{tab:discussion}. We note that the
profile of the Ca {\sc i} 445.48 nm line, which appears in
Figs. \ref{fig:panel-lines} and \ref{fig:vstrips}, and that of the Fe
{\sc ii} 445.16 nm, which also appears in Fig. \ref{fig:vstrips}, are
scaled in those figures to an intensity at the bottom of $\sim\!0.9$ for
plotting purposes. In this figure, all three lines show
their actual relative intensities.}
\label{fig:discussion}
\end{figure}

Figure \ref{fig:discussion} shows, as an illustrative example, the
synthetic spectra of the two sets of models in a short wavelength
interval between 445.0 and 446.0 nm. The upper and lower panels in the
figure correspond to the models in the upper and lower parts of Table
\ref{tab:discussion}.  The spectra have been normalised to the
intensity at 445.2 nm and contain five lines: Fe {\sc ii} 445.16, Ca
{\sc i} 445.48, Fe {\sc ii} 445.58, Ti {\sc ii 445.66}, and Fe {\sc
  ii} 445.91 nm. The colour codes of the spectra — red, black, cyan,
purple, and orange — correspond to decreasing values of $\varv \sin i$
(upper panel) and increasing values of inclinations (lower panel),
the model for Vega being plotted in black.

What is interesting in this plot is how sensitive the profiles are to
changes in inclination and $\varv \sin i$, a conclusion that can be
extended to the full spectral range. In particular, the profiles of
the lines with peculiar shapes, as in the cases of Ca {\sc i}
445.48, Fe {\sc ii} 445.58, and Fe {\sc ii} 445.91 nm, change very
dramatically as the inclination decreases. Very interesting, too, is
the comparison between the behaviour of the normal rounded-shape Fe
{\sc ii} 445.16 nm line and the peculiar Ca {\sc i} 445.48 nm profile
in the upper panel: whereas the Fe {\sc ii} line behaves as one would
expect as the value of $\varv \sin i$ increases, the shape and depth
of the Ca {\sc i} line changes drastically. 

The model plotted as a dotted grey line in the lower panel 
has been computed for $i\!=\!5^\circ$, $\omega\!=\!0.632$, assuming 
the von Zeipel approximation; this model must be compared with 
the one plotted in red, computed with the same parameters, but 
under the assumptions of the $\omega$-model. That value of 
$\omega$ would be associated with a $\beta$ exponent $\sim0.197$, quite
far from $\beta\!=\!0.25$. As can be inferred from the values of 
$T_{\rm eq}/T_{\rm pole}$ for both models in Table \ref{tab:discussion},
the equator is almost 400 K cooler when the von Zeipel approximation 
is used, which results in deeper lines, in particular those with 
peculiar profiles, leading to erroneous determinations of abundances. 
This is a good example of the influence of the $\beta$ exponent 
on the line shapes and intensities.

All this shows the potential of a detailed spectral analysis to find
structural and physical parameters and inclinations of this kind of
stars. Good examples are the works by \citet{takeda2008b},
\citet{Takeda2020}, and \citet{takeda2021} disentangling Sirius A and
Vega's properties using spectral line profiles, or Fourier analysis.

A quantitative analysis of the usefulness of the proposed formalisms
is relevant. Regarding the inclination of the star, it is apparent
that very clear changes are observed in certain line profiles as $i$
moves in the range between 0 and $\sim\!20$ degrees; at larger
inclinations, most of the lines are insensitive to this parameter. It
is easy to prove that the probability of a star having an inclination
between $i$ and $i+\Delta i$ is $P(i,i+\Delta i)\!=\!\sin i\,\Delta
i$, and therefore the probability of finding a star, among a large set of
objects with an inclination in the interval of [0,20] degrees is
$\sim\!0.0603$. As a first impression, one might consider the whole
modelling effort to be disproportionate considering that the number is
small; however, a query to the {\em Gaia} DR3 catalogue asking how
many stars with spectral types between A0 and A9 — for which the
methods presented here would be useful — with parallaxes, $\varpi$, with
relative errors of $\Delta\varpi/\varpi\!<\!0.20$, are $\sim\!146\,000$
($\varpi\geq 2$ mas) and $\sim\!231\,200$ ($\varpi\geq 1$ mas). In other words,
in a sphere with a radius of 1 kpc, we would find around $\sim\!14000$
stars in that range of spectral types with inclinations less than 20
degrees.  The constraint of $-0.037\leq BP-RP \leq+0.377$ to bracket the
interval A0-A9 has been
used.\footnote{https://www.pas.rochester.edu/$\sim$emamajek/}

\section{Conclusions}
\label{sec:conclusions}
In this paper, we provide a combined method to compute the structure of
rapidly rotating stars and build their synthetic spectra. A summary of
the main features of the whole formalism follows:

\begin{enumerate}
    \item The $\omega$-model by \citet{espinosa2011} has been implemented to
      compute the relevant parameters of the photosphere of rapidly
      rotating stars — namely, the radius, $R$, effective temperature,
      $T_{\rm eff}$, and effective gravity, $g_{\rm eff}$ — as a
      function of the colatitude, $\theta$. The method, relatively
      simple from a computational point of view, is able to reproduce
      the results of more complex models. One of the big advantages of
      this formalism is that it avoids the discussion, and hence the
      subsequent computation, or ad hoc assignment, of the appropriate
      gravity darkening exponent, $\beta$. The model is applicable to
      stars with radiative envelopes (Sect. \ref{sec:structure}). 
      In those situations in which some of the approximations 
      inherent in the $\omega$-model — that is, mass concentrated near the
      centre of the star and rigid rotation (Roche model) — are no longer 
      valid, the original ESTER model should be used.

    \item A detailed method of how to compute the synthetic spectrum
      of a rapidly rotating star, at any inclination angle, $i$, with
      respect to the line of sight, is presented. The model makes use
      of the suite of codes, {\sc atlas} and {\sc synthe}
      \citep{kurucz2014}, and the grid of model atmospheres by
      \citet{castelli2003} (Sect. \ref{sec:synthesis} and Appendices
      \ref{app:geometry} and \ref{app:limbdarkening}).
    
    \item The combined methods summarised in items 1 and 2 above were
      applied to the particular case of Vega, obtaining results
      regarding both the structure and the synthetic spectrum that are compatible
      with previous works. The fitting of the spectral lines was
      remarkable, both for those with normal, rounded shapes and those
      with peculiar profiles (Sect. \ref{sec:vega}).

    \item In addition, Appendix \ref{app:geometry} describes in detail
      how to treat, from a strict geometrical point of view, all the
      relevant variables when a rotating star is seen with a given
      inclination with respect to the line of sight.

\end{enumerate}

Although this work has focused on the spectral synthesis of rapid
rotators, the tools provided in this paper can be useful in other
contexts:

\begin{enumerate}
    \item To locate the position of a star in colour-magnitude
      diagrams, since a star deformed by rapid rotation appears
      brighter and hotter when it is observed near pole-on \citep[see
        e.g.][]{perez1999,bastian2009,girardi2019}.

    \item To find and estimate the inclination of the rotation
      axis with respect to the line of sight in those cases without
      making use, in the first instance, of interferometric measurements 
      \citep{takeda2008b}. This would be useful to search for potential 
      pre-main-sequence stars of spectral types earlier than F hosting 
      Jupiter-like planets. In particular, there is indirect evidence 
      that Herbig Ae/Be stars with low metallicities could be good 
      candidates to host such giant planets \citep{kama2015,guzmandiaz2023}. 
      The method presented here would allow a detailed metallicity analysis, 
      and a subsequent filtering of targets according to their inclination, 
      which is suitable in the case of low-inclination systems of potential 
      interferometric and/or direct imaging studies.

    \item The role of the inclination is particularly important in modelling
    accretion processes for young objects of intermediate mass. In the scenario 
    of magnetospheric accretion, the shape and intensity of the spectral lines are 
    strongly dependent on the assumed inclination \citep{muzerolle2004,mendigutia2011}. 
    Regarding the alternative scenario of boundary layer continuum models, the 
    dependence on the inclination is also critical \citep[see e.g. Fig. 5 in][]{mendigutia2020}.
\end{enumerate}
    
\begin{acknowledgements}
The author is very grateful to the referee, Prof. Michel Rieutord, and 
his colleagues, Alain Hui-Bon-Hoa and Axel Lazzarotto, for providing very useful
comments, suggestions and references that, no doubt about, have 
improved the contents and scope of the paper.
This research has been funded by grants AYA2014-55840-P,
PGC2018-101950-B-I00 and PID2021-127289-NB-I00 by the Spanish Ministry
of Science and Innovation/State Agency of Research (MCIN/AEI). The
author is grateful to Francisco Espinosa-Lara for useful discussions
on the ER11 formalism, Antonio Claret for some guiding for the
computation of the limb-darkening coefficients, and Almudena
Alonso-Herrero, Olga Balsalobre-Ruza, Carlos Eiroa, Jorge Lillo-Box,
Ignacio Mendigut\'{\i}a, Enrique Solano and Eva Villaver for their help 
and comments to several sections of this paper. Special thanks also to
Antonio Parras and Sergio Su\'arez for their work keeping up and running 
the computing centre.
\end{acknowledgements}

\bibliographystyle{aa}
\bibliography{oblatestars.bib}

\begin{appendix}
\section{The geometry of the problem}
\label{app:geometry}

According to the notation in Fig. \ref{fig:geometry_1}, a point on the
stellar surface with coordinates

\begin{equation}
\boldsymbol{r}=\left\{
\begin{array}{lr}
x=r\sin\theta\sin\phi  \\
y=r\cos\theta          \\
z=r\sin\theta\cos\phi
\end{array}\right.
\nonumber
\end{equation}

\noindent when the star is seen equator-on, is transformed after a
counterclockwise rotation around the $x$ axis by an angle, $\alpha$, is
done, into $\boldsymbol{r'}\!=\!(x',y',z')$ by applying to
$\boldsymbol{r}$ the matrix $\boldsymbol{\rm R}_x(\alpha)$; namely,
\begin{equation}
\boldsymbol{\rm R}_x(\alpha)=
\begin{pmatrix}
1 & 0          & 0             \\
0 & \cos\alpha & -\sin\alpha   \\
0 & \sin\alpha &  \cos\alpha    
\end{pmatrix}
\nonumber
\end{equation}
\begin{equation}
\boldsymbol{r'}=(x',y',z')=\boldsymbol{\rm R}_x(\alpha)\,\boldsymbol{r}
.\end{equation}

\begin{figure}[!ht]
\centering
\includegraphics[width=9.0cm]{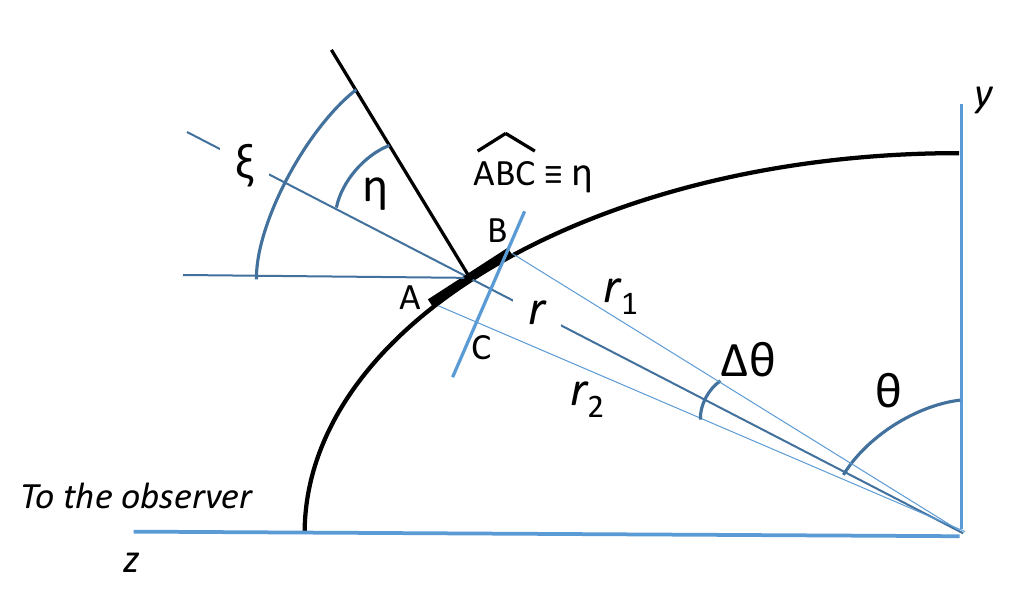}
\caption{Sketch showing the geometrical problem involved in the
  computation of the angles, $\eta$ and $\xi$, as functions of the
  colatitude and the radius of a given surface element.}
\label{fig:geometry_2}
\end{figure}

The observer would see all the points on the stellar surface
fulfilling the easy constraint, $z'= y\sin\alpha + z\cos\alpha > 0$. In
the case of an spherical object, the unitary vector, $\boldsymbol{u}$,
attached to any point has the direction of the normal to the surface;
however, in an oblate object this is not the case, as can be seen
in Fig. \ref{fig:geometry_2}. The normal to the surface at a given
point with colatitude $\theta$, is inclined at an angle of
$\xi\!=\!\pi/2\!-\!\theta\!+\!\eta$ with respect to the line of sight
before proceeding to apply the rotation by an angle, $\alpha$. The
computation of $\eta$ is a fairly straightforward geometrical problem,
as is illustrated in the figure, in which $\Delta\theta$ has obviously been 
plotted out of scale:
$\tan\eta\simeq(r_2-r_1)/(r\,\Delta\theta)$.

A differential surface area element at a latitude, $\theta$, can be
written as $\Delta
A\simeq(r\,\Delta\theta/\cos\eta)\cdot(r\sin\theta\,\Delta\phi)$, and
its associated unitary vector normal to the surface, before the star
is rotated, has the following expression:
\begin{equation}
\boldsymbol{u}_A=(\cos\xi\sin\phi,\sin\xi,\cos\xi\cos\phi)
.\end{equation}

\noindent Therefore, after applying the rotation, the projected area
as seen by the observer would be $\Delta A$ multiplied by the
z-component of $\boldsymbol{\rm R}_x(\alpha)\cdot\boldsymbol{u}_A$;
namely,
\begin{equation}
(\Delta A)_{\rm p} = \Delta A(\sin\alpha\sin\xi+\cos\alpha\cos\xi\cos\phi)
\label{eqn:ap}
.\end{equation}

Concerning the rotation speed of each surface element,
$\boldsymbol{v}$, it has only components $x$ and $z$, as can be seen
in Fig. \ref{fig:geometry_1} (left):

\begin{equation}
\boldsymbol{\varv}=\left\{
\begin{array}{lr}
\varv_x=\Omega\,r\sin\theta\cos\phi \\
\varv_y=0                         \\
\varv_z=-\Omega\,r\sin\theta\sin\phi
\end{array}\right.
\nonumber
.\end{equation}

After applying the rotation, the component of the velocity in the line of sight is
\begin{equation}
\varv'_z=-\Omega\,r \cos\alpha\sin\theta\sin\phi
\label{eqn:vz}
.\end{equation}

Finally, the knowledge of the angle $\gamma$, between the normal to a
given surface element and the line of sight, in the rotated system,
must be known in order to apply the correction for limb darkening to
the synthetic spectra arising from that surface element.

\begin{figure}[!h]
\centering
\includegraphics[width=8.5cm]{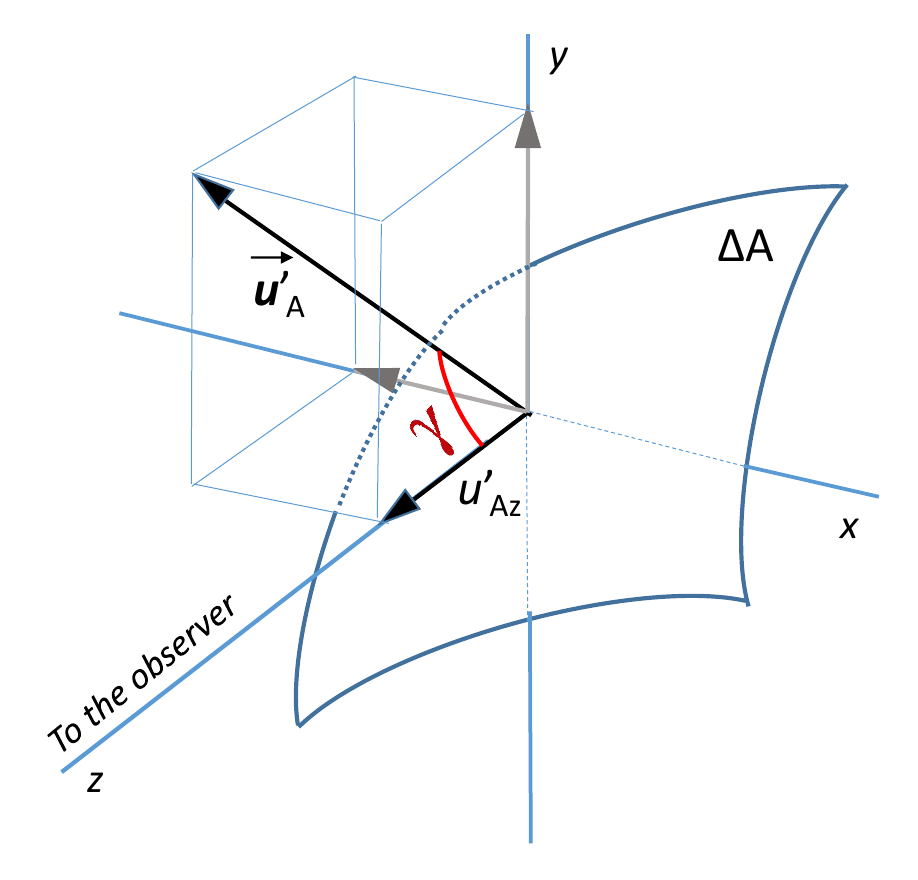}
\caption{Sketch showing the the geometry of a surface element,
  characterised by the unitary vector, $\boldsymbol{u'}_{\rm A}$, whose
  $z$ component forms an angle $\gamma$, with the line of sight.
  $\mu\!=\!\cos\gamma$ is required to compute the limb-darkening
  coefficient (see Appendix \ref{app:limbdarkening})}.
\label{fig:geometry_3}
\end{figure}

In the usual notation for that angle, $\mu\!=\!\cos\gamma$, and
according to Fig. \ref{fig:geometry_3}, it can be written as
$\mu\!=\!\cos (u'_{{\rm A}z}/|\boldsymbol{u'}_{\rm A}|)$, and since
$|\boldsymbol{u'}_{\rm A}|\!=\!1$, the value of $\cos\gamma$ is just
the $z$ component of $\boldsymbol{\rm
  R}_x(\alpha)\cdot\boldsymbol{u}_A$; namely,
\begin{equation}
\mu=\cos\gamma=\sin\alpha\sin\xi+\cos\alpha\cos\xi\cos\phi 
\label{eqn:mu}
.\end{equation}

\section{The limb-darkening coefficients}
\label{app:limbdarkening}

This appendix shows how the limb-darkening coefficients (LDCs
hereafter) and the limb-darkening correction $C_{\rm ld}(\lambda)$
(see eqn. (\ref{eqn:flux}) are computed.  The work and notation by
\citet{claret2011} are followed in this section. The LCDs $a_k, k=1,4$
are defined in such a way that the most general law is adjusted by the
following expression:
\begin{equation}
\frac{I(\mu)}{I(1)}=1-\sum_{k=1}^{4} a_k\,(1-\mu^{k/2})
\label{eqn:ldc}
.\end{equation}
\noindent $I(1)$ is the intensity at the centre of the disc, and
$\mu=\cos\gamma$ (see eqn. (\ref{eqn:mu})), where $\gamma$ is the
angle between the normal to the surface area element and the line of
sight.  Equation (\ref{eqn:ldc}) can be defined for the intensity in a
given passband, although in our case we are interested in a
monochromatic estimate of that quantity for each of the wavelengths
covered by the synthetic spectra.

\citet{claret2011} computed, among others, the LCDs for the photometric
Johnson-Cousins $UBVRI$ filters (Table 17 available at
\citealt{claret2011cat}). In order to estimate the limb-darkening
correction at each wavelength, $\lambda$, we linearly interpolate
the LCDs in this way:
\begin{equation}
a_k(\lambda)= a_k(\lambda_{\rm F1}) + \left[\frac{a_k(\lambda_{\rm F2})-a_k(\lambda_{\rm F1})}
{\lambda_{\rm F2}-\lambda_{\rm F1}}\right]\,(\lambda - \lambda_{\rm F1})
,\end{equation}
\noindent where F1 and F2 stand for ‘Filter 1’ and ‘Filter 2’ and
$\lambda_{\rm F1}$, $\lambda_{\rm F2}$ are the effective wavelengths
of the filters adjacent to the wavelength, $\lambda$, under
consideration; that is, $\lambda_{\rm F1} < \lambda \leq \lambda_{\rm
  F2}$. The wavelengths assigned to the $UBVRI$ filters for our
computations are 360, 440, 550, 690, and 950 nm,
respectively. Following that notation, the correction for
limb-darkening applied to the fluxes at a given wavelength is
\begin{equation}
C_{\rm ld}(\lambda) =  1-\sum_{k=1}^{4} a_k(\lambda)\,(1-\mu^{k/2})
\label{eqn:lcdlambda}
.\end{equation}

\end{appendix}

\end{document}